\documentclass[traditabstract]{aa}
\usepackage{graphicx}
\usepackage{txfonts}
\usepackage[authoryear]{natbib}
\begin{document}

\title{Probing the Astrophysics of Cluster Outskirts}

\author{A. Lapi\inst{1,2}, R. Fusco-Femiano\inst{3}, A. Cavaliere\inst{1}}
\institute{$^1$ Dip. Fisica, Univ. `Tor Vergata', Via Ricerca Scientifica 1, I-00133 Roma, Italy.\\
$^2$ SISSA/ISAS, Via Beirut 2-4, I-34151 Trieste, Italy.\\
$^3$ INAF-IASF, Via Fosso del Cavaliere, 00133 Roma, Italy.}

\date{\today}

\abstract{In galaxy clusters the entropy distribution of the
IntraCluster Plasma modulates the latter's equilibrium within the
Dark Matter gravitational wells, as rendered by our Supermodel. We
argue the entropy production at the boundary shocks to be reduced or
terminated as the accretion rates of DM and intergalactic gas peter
out; this behavior is enforced by the slowdown in the outskirt
development at late times, when the Dark Energy dominates the
cosmology while the outer wings of the initial perturbation drive
the growth. In such conditions, we predict the ICP temperature
profiles to steepen into the cluster outskirts. The detailed
expectations from our simple formalism agree with the X-ray data
concerning five clusters whose temperature profiles have been
recently measured out to the virial radius. We predict steep
temperature declines to prevail in clusters at low $z$, tempered
only by rich environs including adjacent filamentary structures.}

\keywords{galaxies: clusters: general --- galaxies: clusters:
individual (Abell 1795, PKS 0745-191, Abell 2204, Abell 1413,
Abell 1689) --- X-rays: galaxies: clusters --- methods:
analytical.}

\authorrunning{A. Lapi et al.}
\titlerunning{Astrophysics of Cluster Outskirts}

\maketitle

\section{Introduction}

Galaxy clusters constitute the largest bound structures in the
Universe, with their masses up to $M\sim 10^{15}\, M_{\odot}$
and outskirts extending out to sizes $R\sim$ a few Mpcs. These
set the \emph{interface} between the intergalactic environment
keyed to the cosmology at large, and the confined intracluster
plasma (ICP). The latter pervades the clusters at temperatures
$k_B T\propto GM/R\sim 5$ keV and number densities $n\sim
10^{-3}$ cm$^{-3}$, and so emits copious X-ray powers mainly by
thermal Bremsstrahlung \citep[see][]{Sarazin1988}. The ICP
coexists with the gravitationally dominant dark matter (DM)
component in the baryonic fraction $m/M$ close to the cosmic
value $0.16$, and the two build up together from accretion
across the cluster boundary.

The build up comprises an early collapse of the cluster body,
tailing off into a secular development of the outskirts by
smooth accretion and minor mergers \citep[see][]{Zhao2003,
Diemand2007, Vass2008, Navarro2010}. In radius, the body ranges
out to $r\sim r_{-2}$ where the slope of the DM density run
$n(r)$ equals $-2$; the adjoining outskirts extend out to the
current virial radius $R$ with steepening density.

In time, the transition is marked by the redshift $z_t$;
thereafter $r_{-2}$ stays put while $R$ grows larger in a
quasi-static, self-gravitating DM equilibrium \citep[described
through the Jeans equation, see ][]{Lapi2009a}, to imply for
the standard concentration parameter $c\equiv R/r_{-2}$
observed values $c\approx 3.5\, H(z_t)/H(z_{\rm obs})$ in terms
of the Hubble parameter $H(z)$. In the following we adopt the
standard flat cosmology \citep[see][]{Dunkley2009}. So values
of $c$ ranging from $3$ to $10$ correspond for $z_{\rm
obs}\approx 0-0.2$ to \emph{young} or to \emph{old} dynamical
cluster ages $z_t\sim 0.2-3$. The concentration can be directly
if laboriously probed with gravitational lensing
\citep[see][]{Broadhurst2008, Lapi2009b}.

Secular accretion of DM goes along with inflow of intergalactic
gas. The ensuing ICP equilibrium is amenable to the powerful
yet simple description provided by the
Supermodel\footnote{\textsl{IDL} and \textsl{FORTRAN}
algorithms which implement the Supermodel and run in a fraction
of a second on a standard laptop, can be found at
\textsl{http://people.sissa.it/$\sim$lapi/Supermodel/}.}
\citep[SM; see][hereafter CLFF09]{Cavaliere2009}.

Clearly, inflows into the cluster outskirts are
exposed to the cosmological grip. This is the focus of the
present paper.

\section{Entropy run vs. cluster buildup}

The SM expresses in full the hydrostatic equilibrium (HE) of
the ICP in terms of DM gravity and of the `entropy' $k\equiv
k_BT/n^{2/3}$. In its basic form, the latter's \emph{physical}
run may be represented as
\begin{equation}
k(r)=k_c+(k_R-k_c)\,(r/R)^a~,
\end{equation}
consistent out to $r\approx R/2$ with recent analyses of wide
cluster samples \citep{Cavagnolo2009, Pratt2010}. This embodies
two \emph{specific} ICP parameters: the central level $k_c$ and
the outer powerlaw slope $a$.

The former is set at a basal level $k_c\sim 10$ keV cm$^2$ by
intermittent entropy injections by central AGN feedback
\citep[e.g.,][]{Cavaliere2002, Valageas1999, Wu2000,
McNamara2007}; the ensuing quasi-stable condition corresponds
to cool core morphologies \citep[CC; see][]{Molendi2001},
featuring a limited central temperature dip and generally large
concentrations $c\approx 6-10$, as discussed by CLFF09.

On the other hand, $k_c$ may be enhanced up to several $10^2$
keV cm$^2$ by deep mergers \citep[e.g.,][]{McCarthy2007,
Markevitch2007}, frequent during the cluster youth; these
events give rise to non cool core (NCC) clusters, featuring
generally low concentrations $c\approx 3-5$ and a central
temperature plateau, scarred in some instances by imprints from
recently stalled blastwaves \citep[see discussions
by][]{Fusco2009, Rossetti2010}.

\subsection{The outer regions}

The second term in Eq.~(1) describes the powerlaw outward rise
expected from the scale-free stratification of the entropy
continuously produced by the boundary accretion shock, while
the cluster grows larger by slow accretion.

The slope $a_R$ at $r\approx R$ with standard values around $1$
has been derived by CLFF09 from the shock jumps and the
adjoining HE maintained by thermal pressure, to read
\begin{equation}
a_R = 2.37-0.47\, b_R~.
\end{equation}
Here $b_R\equiv \mu m_p\,v_R^2/ k_B T_R$ marks the ratio of the
potential to the thermal energy of the ICP
\citep[see][]{Lapi2005, Voit2005}. This reads $b_R\approx
3\,v_R^2/2\,\Delta\Phi$ when a strong shock efficiently
thermalizes into $3$ degrees of freedom the infall energy; the
latter is expressed in terms of the potential drop
$\Delta\Phi=-\int_{R_{\rm ta}}^{R}{\mathrm{d}r}~G\,\delta
M/r^2$, experienced by successive shells of DM and gas that
expand; owing to the excess mass $\delta M$ they turn around at
the radius $R_{\rm ta}\approx 2\,R$ to start their infall
toward the shock at $R$.

At any given cosmic time $t$, Eq.~(2) holds at the current
virial radius $R(t)$. On the other hand, the entropy deposited
there is conserved during subsequent compressions of the
accreted plasma into the DM gravitational well, while no other
major sources or sinks of entropy occur down to the central
$10^2$ kpc. Thus while the cluster outskirts develop out to the
current radius $R(t)$, the specific entropy stratifies with a
running slope $a(r)=a_{R(t)}$ that retains the sequence of
original values set at the times of deposition
\citep[see][]{Tozzi2001}.

Values $a\approx 1$ obtain on adopting the standard ratio
$R/R_{\rm ta}$ $\approx 0.5$, and the simple potential drop
$\Delta \Phi/v_R^2\approx 1-(R/R_{\rm ta})\approx 0.5$
associated to a flat initial mass perturbation $\delta
M/M\propto M^{-\epsilon}$ with $\epsilon\sim 1$ that describes
the collapse of the cluster \emph{body} as a whole. This
implies $b_R\approx 3$ and $a\approx 1$ from Eq.~(2). In fact,
$\Delta\Phi/v_R^2\approx 0.57$ obtains from the full DM
$\alpha$-profile (see CLFF09), implying $b_R\approx 2.7$ and
the standard value $a\approx 1.1$.

However, considerable variations of $a(r)$ are to be expected
in the cluster outskirts, as discussed below.

\begin{figure}
\centering
\includegraphics[width=9cm]{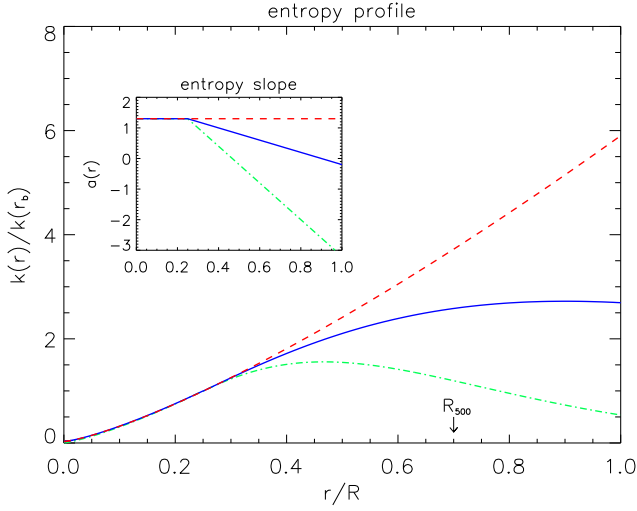}
\caption{Examples of entropy profiles (normalized
at $r=r_b$) and slopes (inset). Dotted line:
after Eq.~(1) with $k_c=0$ and $a=1.3$. Solid line:
Eq.~(4) with $k_c=0$, $a=1.3$, $r_b=0.25\,R$, and $a'=0.5$.
Dashed line: Eq.~(4) with $k_c=0$, $a=1.3$, $r_b=0.25\,R$, and
$a'=1.5$.}
\end{figure}

\subsection{Development of the outskirts}

The cluster \emph{outskirts} for $r>r_{-2}$ originate from the
wings of a realistically bell-shaped perturbation, that may be
described by $\delta M/M\propto M^{-\epsilon}$ for $\epsilon$
exceeding $1$ \citep[e.g.,][]{Lu2006}. Then the outer potential
drop
\begin{equation}
{\Delta\Phi\over v_R^2}={1-(R/R_{\rm
ta})^{3\epsilon-2}\over 3\epsilon-2}~
\end{equation}
is shallower relative to the body value, so leading to higher
values of $b_R$ and \emph{lower} values of $a$. Thus as $r$
increases outwards we expect $k(r)$ to deviate downward from a
simple powerlaw.

The argument may be phrased in terms of the accretion rate
$\dot M$. A shell $\delta M$ enclosing the mass $M$ will
collapse when $\delta M/M$ attains the critical threshold
$1.686\, D^{-1}(t)$ in terms of the linear growth factor $D(t)$
\citep[e.g.,][]{Weinberg2008}. So the shape parameter
$\epsilon$ also governs the mass buildup after $M\propto
D^{1/\epsilon}\propto t^{d/\epsilon}$; here we have represented
the growth factor as $D(t)\propto t^{d}$ with $d$ ranging from
$2/3$ for $z\ga 1$ to approach $1/2$ as $z\rightarrow 0$. So
the outskirts develop from the inside-out, at accretion rates
$\dot M/M\approx d/\epsilon\,t$ that \emph{lower} for
$\epsilon$ exceeding $1$, and for $d$ decreasing toward $1/2$
at late cosmic times\footnote{Note that $d\rightarrow 0$ would
occur if the Dark Energy density increased with time to cause
an ultimate Cosmic Doomsday \citep[see][]{Caldwell2003}; this
would imply truly vanishing accretion rates, and result into a
cutoff of the temperature profiles.}. We add that at small
accretion rates the shock position outgrows $R$
\citep[see][]{Voit2003}, while the shock strength may weaken;
both these effects will decrease $a$ relative to Eqs.~(2) and
(3), and will be discussed in detail elsewhere.

So we see that \emph{flatter} slopes $a$ of the entropy are to
prevail for \emph{decreasing} accretion rates $\dot M$ of DM
and gas; these have a twofold origin. First, the cosmological
structure growth slows down at later cosmic times (low $z_{\rm
obs}$), as expressed by $d<2/3$. Second, perturbation wings
marked by $\epsilon>1$ imply shallow gravitational wells and
little available mass to accrete in average environs; the
effect may be locally offset (and represented with a smaller
effective $\epsilon$) in specifically rich environments,
including adjacent filamentary large-scale structures.

The decline of $a$ from the body value and the entropy bending
set in at a radius $r_b\approx r_{-2}$ where matter began
stratifying onto the outskirts just after $z_t$. Such a radius
is evaluated in terms of the observed concentration in the form
$r_b/R\approx r_{-2}/R\approx 1/c$, to take on values around
$0.2-0.3$ for typical concentrations $c\approx 6-8$ of CC
clusters. Hints to this trend loom out in the data by
\citet{Pratt2010} and \citet{Hoshino2010}.

To sum up, under the \emph{lower} accretion rates prevailing at
\emph{later} times in average environs, we expect the entropy
run to \emph{flatten} out or even \emph{decline} into the
cluster outskirts; then the temperature will decline as $k_B
T(r)= k\,n^{2/3}\propto r^{-2}$ or steeper, after Eq.~(7) of
CLFF09. Do such behaviors show up in real clusters?

\begin{figure}
\begin{minipage}{\linewidth}
\centering
\includegraphics[width=9cm]{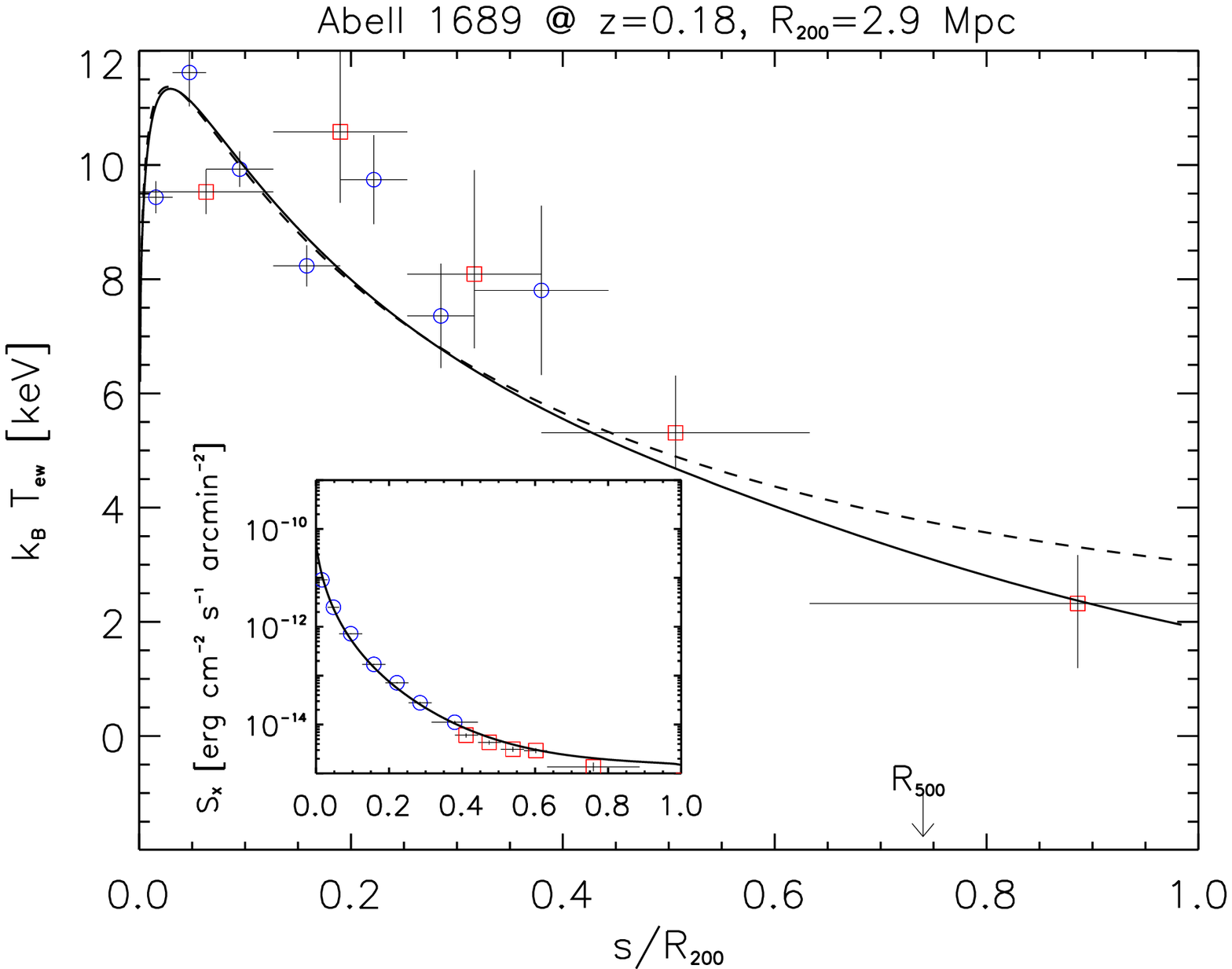}
\vspace{0.5cm}
\end{minipage}
\begin{minipage}{\linewidth}
\centering
\includegraphics[width=9cm]{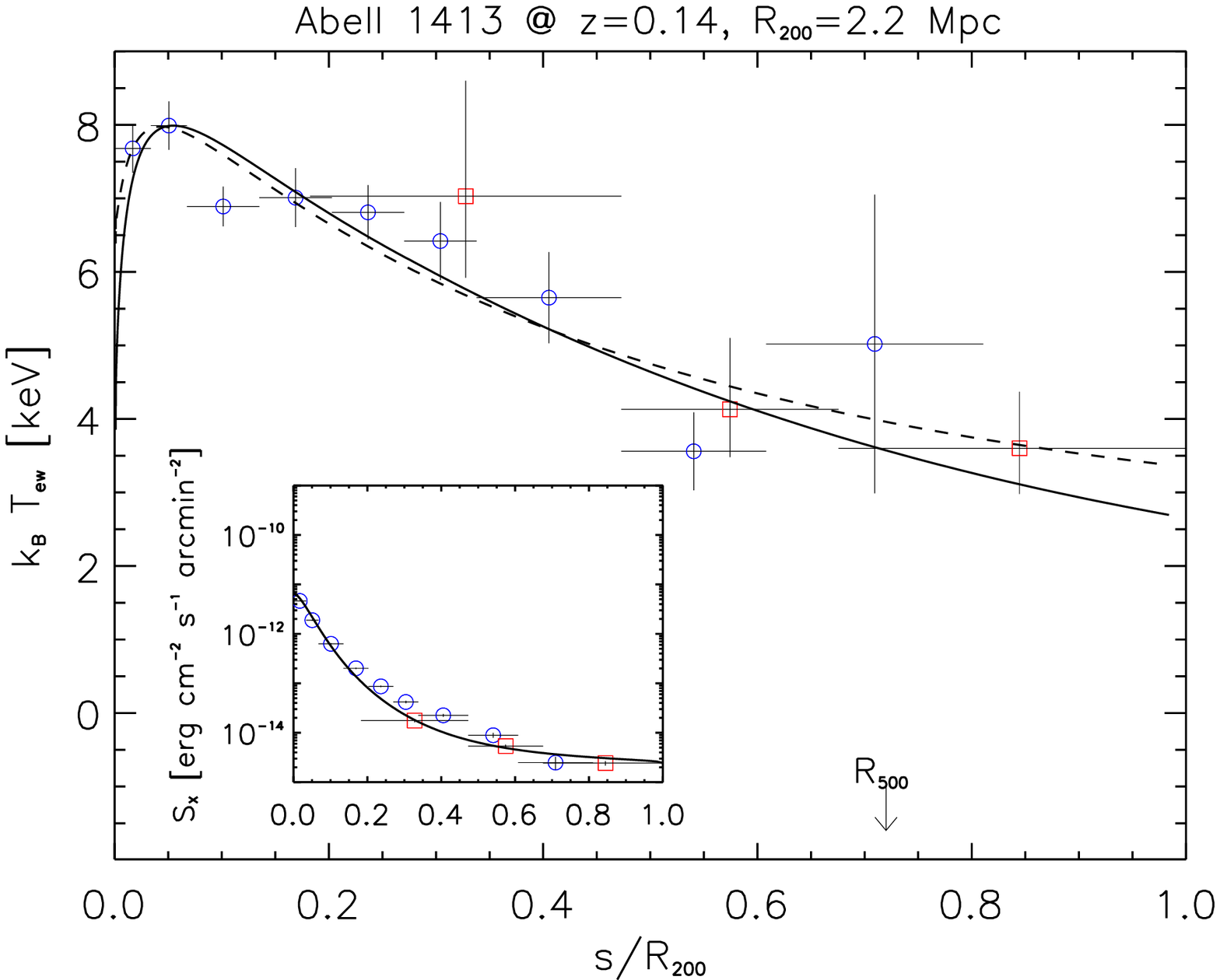}
\vspace{0.5cm}
\end{minipage}
\begin{minipage}{\linewidth}
\centering
\includegraphics[width=9cm]{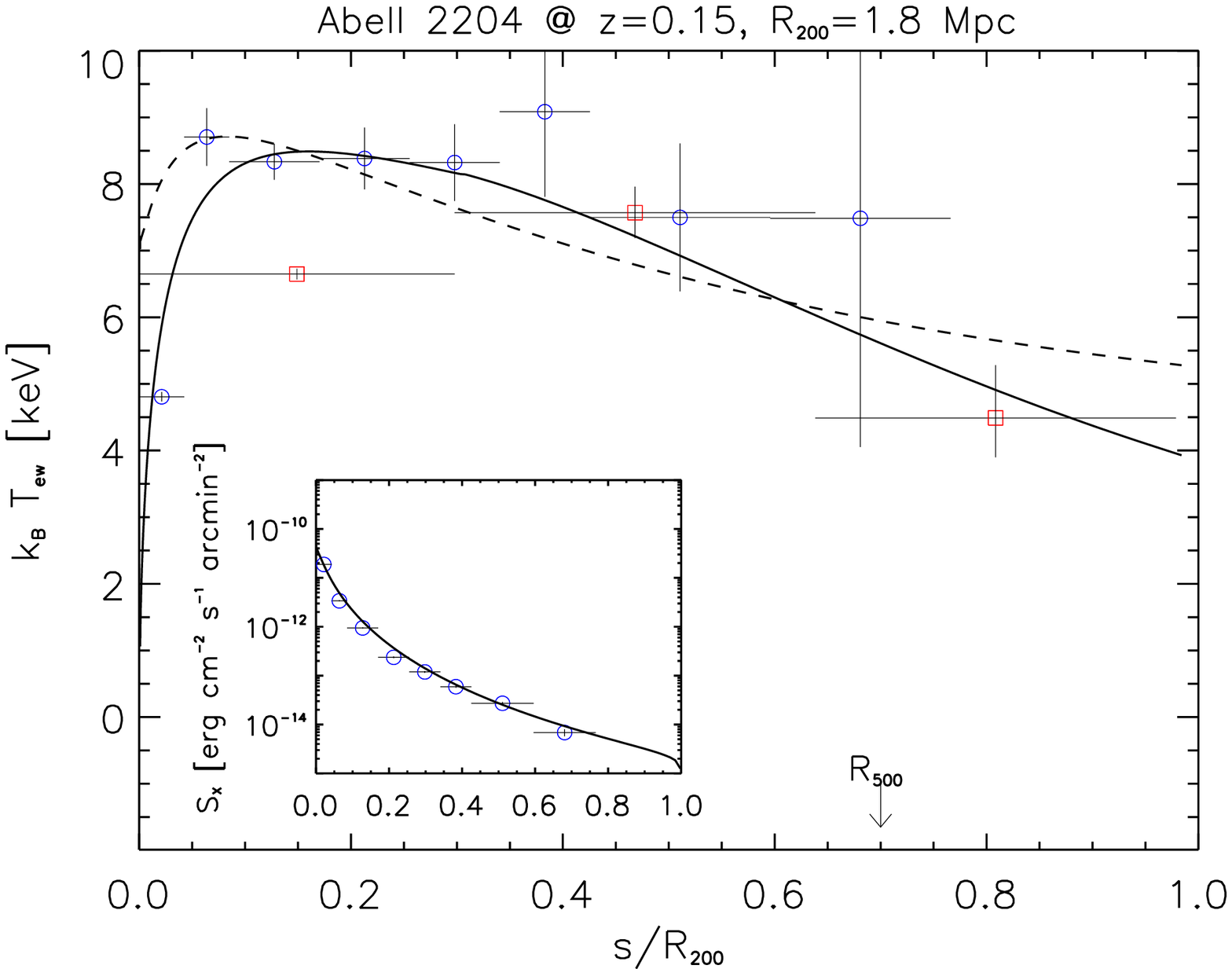}
\vspace{0.5cm}
\end{minipage}
\caption{Profiles of projected X-ray temperature (brightness in
the insets) for the CC clusters A1689 (top), A1413 (middle) and
A2204 (bottom). Data are from \citet{Snowden2008} with
\textsl{XMM-Newton} (blue circles) and from
\citet{Kawaharada2010}, \citet{Hoshino2010}, and
\citet{Reiprich2009} with \textsl{Suzaku} (red squares). Our
best-fits with the SM after Eq.~(4) are illustrated by the
solid lines, while dashed lines refer to the fits based on
Eq.~(1).}
\end{figure}

\begin{figure}
\begin{minipage}{\linewidth}
\centering
\includegraphics[width=9cm]{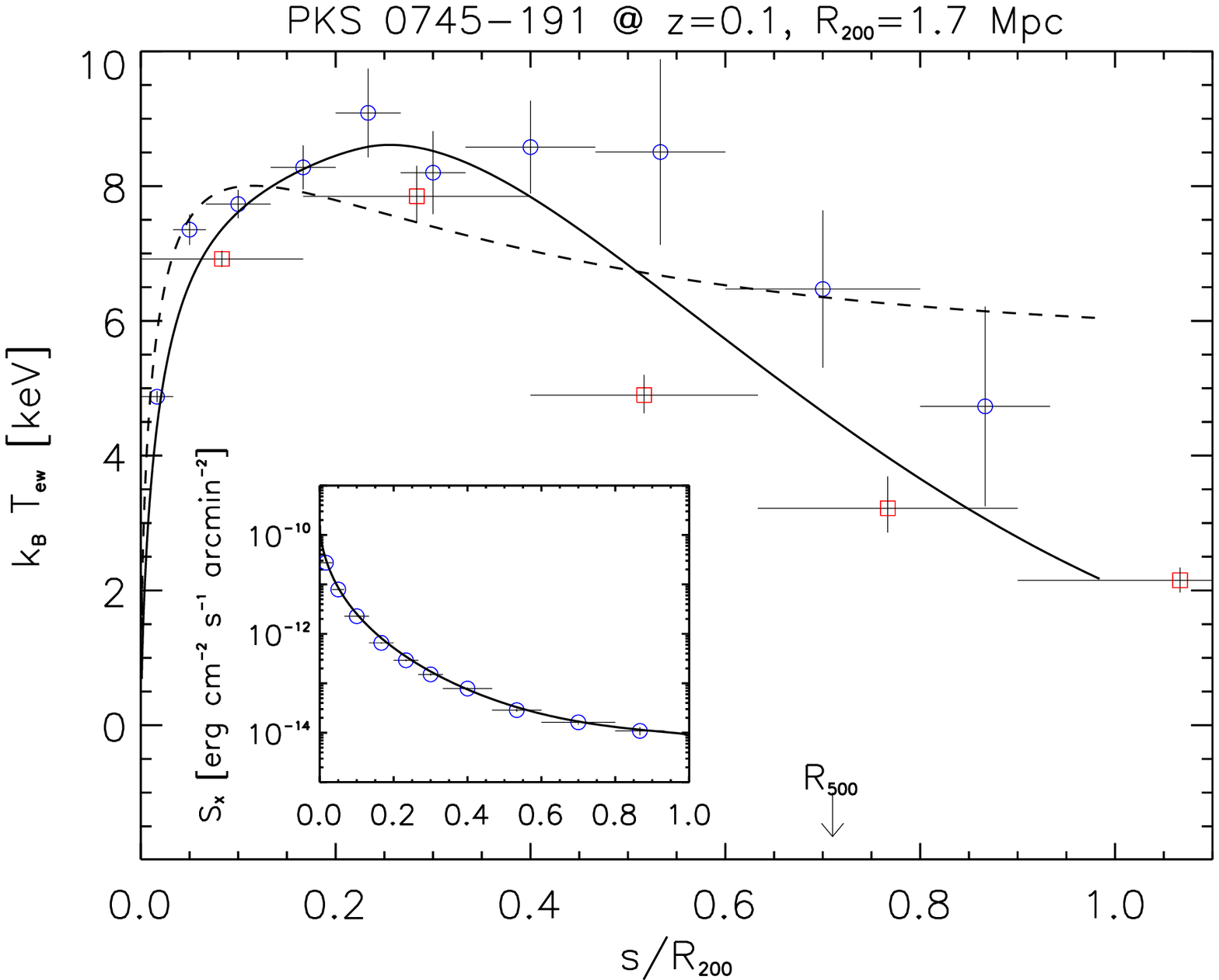}
\vspace{0.5cm}
\end{minipage}
\begin{minipage}{\linewidth}
\centering
\includegraphics[width=9cm]{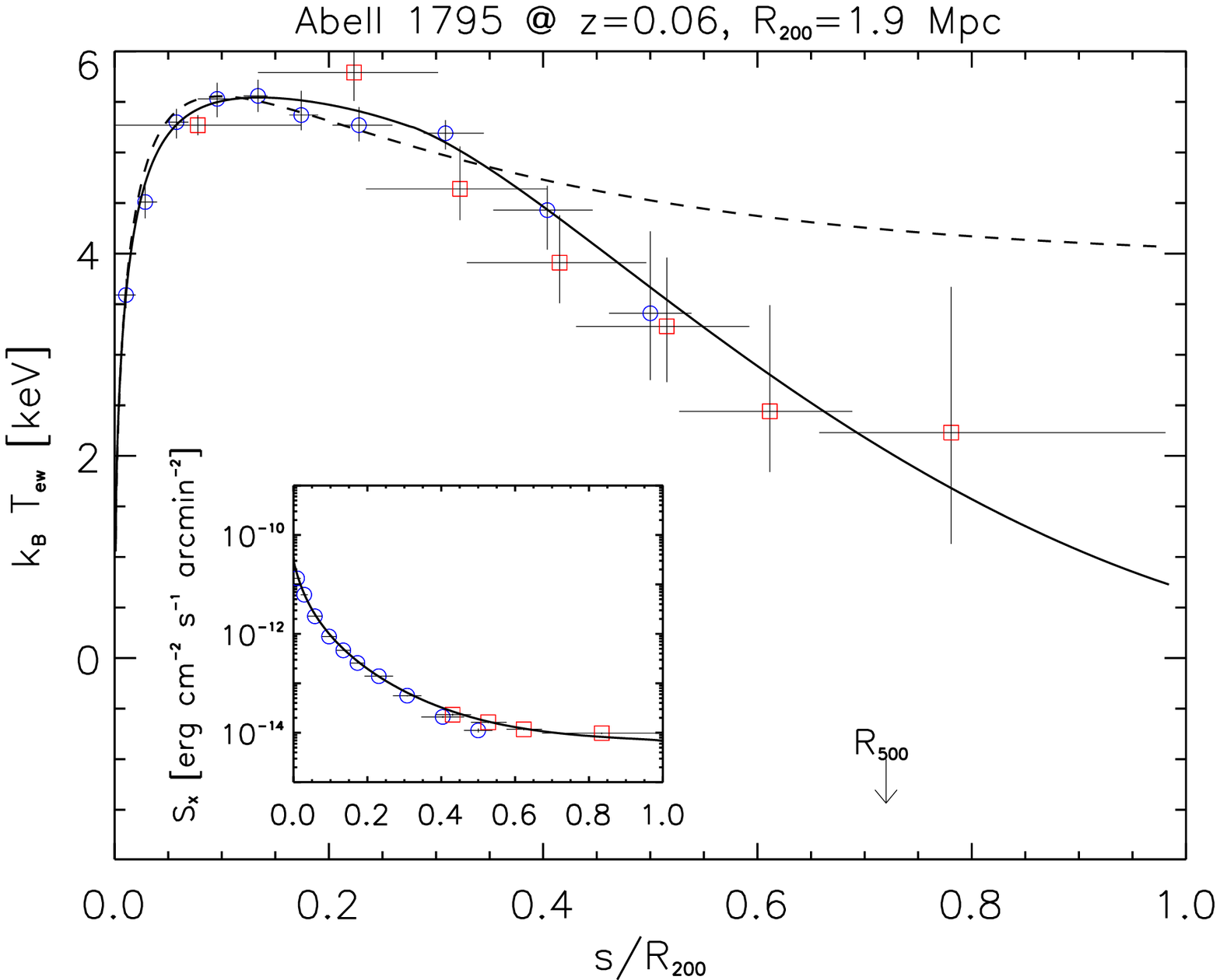}
\vspace{0.5cm}
\end{minipage}
\caption{Same as Fig.~2 for the low-$z$ clusters PKS0745-191 (top) and A1795
(bottom). Data are from \citet{Snowden2008} with
\textsl{XMM-Newton} (blue circles) and from \citet{George2009}
and \citet{Bautz2009} with \textsl{Suzaku} (red squares).}
\end{figure}

\begin{table}
\caption{Fitting parameters from ICP temperature profiles}
\begin{center}
\begin{tabular}{lccccc}
\hline\hline
Cluster & $c$ $^*$ & $a$ & $r_b/R$ & $a'$ & $\chi^2$ $^{**}$ \\
\hline
\\
A1795 & $8.5^{+1.9}_{-1.9}$ & $1.2^{+0.3}_{-0.3}$  & $0.28^{+0.02}_{-0.02}$ & $1.8^{+1.3}_{-1.3}$ & $0.3\, (2.6)$\\
\\
PKS0745-191 & $7.6^{+1.7}_{-1.7}$ & $1.9^{+1.3}_{-1.3}$  & $0.23^{+0.03}_{-0.03}$ & $1.1^{+0.7}_{-0.7}$ & $1.4\, (4.4)$ \\
\\
A2204 & $5.5^{+1.1}_{-1.1}$ & $1.5^{+1.1}_{-1.1}$ & $0.31^{+0.07}_{-0.07}$ & $0.6^{+0.4}_{-0.4}$ & $1.1\, (2.1)$ \\
\\
A1413 & $8.3^{+1.7}_{-1.7}$ & $0.9^{+0.3}_{-0.3}$  & $0.27^{+0.07}_{-0.07}$ & $0.2^{+0.03}_{-0.03}$ & $1.2\, (1.9)$ \\
\\
A1689 $^\dag$ & $12.4^{+5.3}_{-5.3}$ & $0.7^{+0.3}_{-0.3}$  & $0.5^{+0.1}_{-0.1}$ & $1.6^{+1.2}_{-1.2}$ & $1.5\, (1.7)$ \\
\\
\hline
\end{tabular}
\end{center}
$^{*}$ DM concentration estimated from X rays.
\\
$^{**}$ Reduced $\chi^2$ values for fits with the entropy run
in Eq.~(4); in parenthesis the values with $a'=0$,
corresponding to Eq.~(1).
\\
$^\dag$ CC classification is controversial
\citep[see][]{Riemer2009}.
\end{table}

\section{A case study on current data}

Toward an answer, we use the SM to provide profiles of density
and temperature from expressing the expected entropy run in a
simple form: the initial slope $a$ still applies in the cluster
body for $r\leq r_b$; but for $r> r_b$ it goes over to a
decline toward the current boundary value $a_R<a$ following
$a-a'\,(r/r_b-1)$ with a constant gradient $a'$. So the entropy
profile reads
\begin{equation}
k(r)=\left\{
  \begin{array}{lll}
    k_c+(k_b-k_c)\,(r/r_b)^a & ~~~r \leq r_b \\
\\
    k_R\, (r/R)^{a+a'}\,e^{a'\,(R-r)/r_b}   & ~~~r> r_b
  \end{array}
\right.~.
\end{equation}
The outer branch describes a linear decline of the slope with
the gradient $a'\equiv (a-a_R)/(R/r_b-1)$; normalizations have
been set as to obtain for $k(r)$ a continuous function and
derivative. We illustrate in Fig.~1 examples of entropy
profiles after Eqs.~(1) and (4), with parameters indicated by
the analyses below.

Adding to the several CCs and NCCs previously analyzed out to
$R/2$ in terms of the standard entropy run \citep[see CLFF09
and ][]{Fusco2009}, here we focus on five CCs with data now
available out to $r\approx R$, and analyze them in terms of the
entropy run of Eq.~(4). We report in Table 1 the resulting
parameters with $68\%$-level uncertainties. For these old CC
clusters we take $k_c\approx 10$ keV cm$^2$ as anticipated in
\S~2.

\section{Discussion and conclusions}

These data clearly bear out our expectations of steep runs for $T(r)$, stemming
from outer entropy production being reduced as the accretion
rates $\dot M$ of DM and intergalactic gas peter out (see
\S~2). We trace back such a behavior to two concurring sources:
(i) the cosmological slowdown in the growth of outskirts
developing at late cosmic times; (ii) shallow perturbation
wings scantily feeding the outskirts growth in average or poor
environs.

Under such conditions, we show that on average the entropy
profiles are to \emph{flatten} out or even \emph{decline} into
the cluster outskirts as represented by Eq.~(4). Then our
Supermodel yields \emph{steep} outer profiles of projected
temperatures (and flatter densities), in close agreement with
current data out to $R$.

We also expect the cosmological decrease of the accretion rates
to be locally offset in rich environs \emph{biased} high, or in
cluster \emph{sectors} adjacent to filamentary large-scale
structures; more standard runs of temperature are predicted
there. These loom out in a sector of A1689 as observed by
\citet{Kawaharada2010}.

Compared with numerical simulations like the ones by
\citet{Nagai2007} and \citet{Roncarelli2006}, our temperature
declines and entropy dearths are comparable as for clusters
like A2204 and A1413; but they are quite sharper for low $z$
clusters like A1795 and PKS0745-191, well beyond possible
contaminations due to \textsl{Suzaku} PSF smearing
\citep[see][]{Reiprich2009}. Low-noise, high-resolution
simulations addressing the issue of entropy production related
to richness of surrounding environment or adjoining filamentary
structures will be most fruitful.

We have checked that with the SM the gas mass $m(<r)$ grows
monotonically outwards, and actually faster than the DM's, so
as to produce an increasing baryonic fraction by a factor $10$
from $R/20$ to $R$ \citep[cf.][]{Zhang2010}. So sharply
declining $T(r)$ and increasing masses are in fact consistent
with thermal HE. Note that as $\dot{M}$ decreases to the point
that the infall velocities decrease to transonic values, the
shocks weaken, thermalization becomes inefficient (see CLFF09),
the entropy production is terminated, and thermal pressure
alone cannot support HE any longer.

Equilibrium in the outskirts can be helped by bulk or
turbulent, merger-induced motions; these contribute up to
$10-20\%$ of the total support in relaxed clusters, as gauged
in terms of pressure and X-ray masses both observationally
\citep[][adding to the refs. in \S~3]{Mahdavi2008, Zhang2010}
and numerically \citep[][]{Nagai2007, Piffaretti2008, Bode2009,
Lau2010, Meneghetti2010}. A similar accuracy is intended for
our predictions based on thermal HE, while a finer
approximation with non-thermal contributions included in the SM
will be provided elsewhere.

On the other hand, even extended HE constitutes just a useful
approximation, due to fail even in relaxed clusters:
locally, when minor lumps of cold gas fall into the cluster from
an adjacent filament, along with smooth accretion; globally, beyond
$R$ where the DM equilibrium is in jeopardy.

In sum, from the SM with thermal HE and optimal energy
conversion in strong shocks, we find temperature profiles
declining \emph{sharply} outwards. These stem from progressive
\emph{exhaustion} of mass inflow, especially for clusters in
average or poor environments at \emph{low} $z_{\rm obs}$; such
a predicted trend is consistent with the current evidence for
the handful of clusters in Figs.~2 and 3.

Our punchline is that cosmology - besides affecting
the cluster statistics like mass or temperature distributions
\citep[see][]{Vikhlinin2009} - concurs with the perturbation
shape to set the outer structure of \emph{individual}
clusters and their development.

All these rich phenomena expected at the \emph{interface}
between the ICP and the intergalactic medium call for extensive
probing even at $z\ga 0.2$ with the next generation of X-ray
telescopes planned to detect at high resolutions low-surface
brightness plasma \citep[see][]{Giacconi2009}, like
\textsl{WFXT} (see \textsl{http://wfxt.pha.jhu.edu/}) and
eventually \textsl{IXO} (see
\textsl{http://ixo.gsfc.nasa.gov/}).

\begin{acknowledgements}
Work supported by ASI and INAF. We thank our referee for useful
comments. AL thanks SISSA and INAF-OATS for warm hospitality.
\end{acknowledgements}

\end{document}